\documentclass[letterpaper, 10 pt, conference]{ieeeconf}  

\usepackage{amsmath}
\usepackage{graphicx} 
\usepackage{enumerate}
\usepackage{mathpazo} 
\usepackage[T1]{fontenc} 
\usepackage{amssymb}
\usepackage[Symbol]{upgreek}
\usepackage{amsmath}
\usepackage{bm}
\usepackage[german]{babel}

\IEEEoverridecommandlockouts                              

\title{\LARGE \bf
Error Analysis for the Particle Filter:\\ Methods and Theoretical Support 
}

\author{Ziyu Liu$^{1}$, Shihong Wei$^{2}$, and James C. Spall$^{3}$
\thanks{A compressed version of this paper appears in the Proceedings of the American Control Conference, Philadelphia, PA, 10-12 July 2019.}
\thanks{$^{1}$Ziyu Liu is a graduate student of Johns Hopkins University Applied Math and Statistics Department. 
        Whitehead Hall, 3400 North Charles Street, Baltimore, MD 21218
        {\tt\small zliu82@jhu.edu}}%
\thanks{$^{2}$Shihong Wei is a graduate student of Johns Hopkins University Applied Math and Statistics Department.
        Whitehead Hall, 3400 North Charles Street, Baltimore, MD 21218
        {\tt\small swei15@jhu.edu}}%
\thanks{$^{3}$James C. Spall is a member of the Principal Professional Staff at the JHU Applied Physics Laboratory and Research Professor of Johns Hopkins University Applied Math and Statistics Department.
        Whitehead Hall, 3400 North Charles Street, Baltimore, MD 21218
        {\tt\small James.Spall@jhuapl.edu}}%
}

\begin{document}

\maketitle
\thispagestyle{empty}
\pagestyle{empty}

\begin{abstract}

The particle filter is a popular Bayesian filtering algorithm for use in cases where the state-space model is nonlinear and/or the random terms (initial state or noises) are non-Gaussian distributed. We study the behavior of the error in the particle filter algorithm as the number of particles gets large. After a decomposition of the error into two terms, we show that the difference between the estimator and the conditional mean is asymptotically normal when the resampling is done at every step in the filtering process. Two nonlinear/non-Gaussian examples are tested to verify this conclusion.

\end{abstract}

\section{INTRODUCTION}

This paper is aimed at error analysis for the particle filter (PF) in the nonlinear and/or non-Gaussian discrete state-space model. We establish the asymptotic normality for the difference between the PF estimate and the conditional mean in multivariate cases.

The PF, proposed in \cite{1993}, is a popular Bayesian filtering algorithm for its ease of implementation and wide range of application. The PF circumvents the intractability of the required integral operations when updating the posterior density by directly approximating the posterior distributions by a large number of particles instead. Recall that under the linear model and Gaussian noise cases, the Kalman filter is the standard filtering choice, which is exactly the conditional mean. Furthermore, the error distribution of the Kalman filter can be characterized by the covariance matrix. However, the same convenience does not naturally hold for PF. If we have a knowledge of the error distribution of PF estimate, the evaluation of PF algorithm can be more precisely made in specific applications. Moreover, it also helps to improve the performance of PF in terms of deciding the optimal number of particles and tuning crucial parameters in PF variants. Thus, the error analysis for PF is a topic of both theoretical and practical interest.

Many previous studies have been conducted on modifying the generic PF to improve its performance under certain circumstances. In addition, some research has focused on the statistical properties of the PF estimate and among them the characterization of the error term draws some attention. Unlike the error behavior of Kalman filter or extended Kalman filter (see e.g. \cite{19992}), which can be characterized by the error covariance matrix in the presence of Gaussian noise, there is usually no closed-form expression for the error in PF, especially in the nonlinear and/or non-Gaussian system cases. 

However, it is possible to calculate an estimation error bound for some cases of the Kalman filter with non-Gaussian noise \cite{A}\cite{B}. Some studies are made on how PF estimate converges to the true conditional mean as an approximation. Ref. \cite{1999} discussed convergence of PF and the fluctuation of its path space and showed that the distribution of PF converges to the distribution of conditional mean as number of particles increase under certain assumptions. Ref. \cite{2015} studied the distance between the PF as a numerical approximation and its underlying continuous system and then established the convergence of PF to the continuous optimal filter. Other researchers directly focus on characterizing the error distribution. In discrete state-space model, the general framework is to decompose the error term into two parts:
\begin{equation}\label{important}
\hat{\bm{x}}_k-\bm{x}_k = (\hat{\bm{x}}_k - E[\bm{x}_k|\bm{z}_{1:k}]) + (E[\bm{x}_k|\bm{z}_{1:k}] - \bm{x}_k ).
\end{equation}
where $\bm{z}_{1:k}$ represent $(\bm{z}_{1}, \cdots, \bm{z}_{k})$, the first part being the difference between the PF estimate $ \hat{\bm{x}}_k $ and the conditional mean $ E[\bm{x}_k|\bm{z}_{1:k}] $, and second part being the difference between the conditional mean $ E[\bm{x}_k|\bm{z}_{1:k}] $ and the underlying true state $ \bm{x}_k $.

For the first part of the error decomposition, \cite{2013} uses the result of \cite{2012} to show that the distribution of the difference between generic PF estimate and the conditional mean is asymptotically normal in scalar cases as the number of particles gets large. Recently, \cite{2018} conducted error analysis specifically on the linear feedback PF, which, as a special variant of PF, includes a feedback control for particles. However, whether a similar result holds for the multivariate case remains unclear and the theoretical foundation for the second part of the error term also remain to be explored. 

In this paper, we provide an error analysis for a generic type of PF for which re-sampling is performed at every step, with a focus on the first part of the error decomposition \eqref{important}. In fact, we will extend the work of \cite{2013} to allow for the analysis of the difference between estimator and conditional mean under the multivariate case. After a rigorous derivation, we show that the first error term will converge asymptotically to the multivariate normal distribution as the number of particles gets large. Then, we verify the above result on two nonlinear and/or non-Gaussian discrete state-space cases. 

The reminder of this paper is arranged as follows: The second section will be the statement of problem setting and clarification of notation. The third section will be the mathematical analysis showing the asymptotic normality for the partial error term. The fourth section will be the numerical verification on two examples and the last section will the conclusion and discussion for future work.

\section{Problem Statement and Particle Filter}

\subsection{Discrete Time-State-Space Model}

Consider the discrete time state-space model (DSSM) with the state equation and the measurement equation as follows:
\begin{equation}\label{eq*}
\left\{
\begin{array}{lr}
\bm{x}_{k+1}=\bm{f}_k(\bm{x}_k,\ \bm{w}_k),   \\
\bm{z}_k=\bm{h}_k(\bm{x}_k,\ \bm{v}_k) .
\end{array}
\right.
\end{equation}
where $ \bm{x}_k$  and $ \bm{z}_k $ are the true states and the measurements, respectively, with $ \bm{w}_k $ and $ \bm{v}_k $ being the noise terms in the state equation and measurement equation, $\bm{f}_k$ is a possibly nonlinear function of state $\bm{x}_{k-1}$, and $\bm{h}_k$ is a possibly nonlinear function of $\bm{x}_{k}$.

Consider $\{\bm{x}_k\} $ as a  hidden Markov process (HMP), $\bm{x}_k \sim p_k(\cdot| \bm{x}_{k-1})$, $\bm{z}_k \sim p_k(\cdot | \bm{x}_k)$. Denote the historical records of true states and measurements by $\bm{x}_{1:k} = (\bm{x}_1, \cdots, \bm{x}_k)$ and $\bm{z}_{1:k} = (\bm{z}_1, \cdots, \bm{z}_k)$. Then, for this filtering problem, the goal is to calculate $ E\left(\bm{x}_k|\bm{z}_{1:k}\right) $. In the nonlinear and/or non-Gaussian cases, Bayesian filter updates its estimators using the following recursive form:\\
\newline
$Prediction$: using information of $\bm{z}_{1:k-1}$ to predict $\bm{x}_k$
\begin{equation}\notag
\begin{split}
p(\bm{x}_k|\bm{z}_{1:k-1})
=\int{p(\bm{x}_k|\bm{x}_{k-1}) p(\bm{x}_{k-1}{|\ \bm{z}}_{1:k-1})d\bm{x}_{k-1}},
\end{split}
\end{equation}
$Update$: using information of $\bm{z}_{k}$ to adjust $\bm{x}_k$
\begin{align}\notag
p(\bm{x}_k|\bm{z}_{1:k})
=\frac{p(\bm{z}_k|\bm{x}_k)p(\bm{x}_k|\bm{z}_{1:k-1})}{p(\bm{z}_k|\bm{z}_{1:k-1})}.
\end{align}

\subsection{Particle Filter}

For the nonlinear and non-Gaussian DSSM, the integration of the posterior density, as required in computing $E(\bm{x}_k|\bm{z}_{1:k})$, is often intractable. Hence there is usually no closed-form solution to $E(\bm{x}_k|\bm{z}_{1:k})$. However, the PF can be used to represent the posterior density by a set of randomly (re)sampled weighted particles generated by the Monte Carlo method, and the particles can be averaged to form an estimator of the expectation of interest. Let the number of particles be $m$ and the $i^{th}$ particle at time $k$ be $\bm{x}^i_k$. The realization of PF relies heavily on the principle of importance sampling. 


Suppose $p(\bm{x}|\bm{z})$ is our target possiblility density function (p.d.f), and $q(\bm{x}|\bm{z})$ is the proposal p.d.f. Then the unnormalized weights are:
\begin{align}\label{eq7}
{\upalpha_{k}}(\bm{x}^i_{1:k})
&=\frac{p(\bm{z}_{1:k}|{\bm{x}^i_k})p({\bm{x}^i_k})}{q({\bm{x}^i_k}|\bm{z}_{1:k})}  \notag \\ 
&= \frac{p(\bm{z}_{1:k}, \bm{x}^i_k)}{q(\bm{x}^i_k|\bm{z}_{1:k})}  = \frac{p({\bm{x}^i_k}|\bm{z}_{1:k})p({\bm{z}_k})}{q({\bm{x}^i_k}|\bm{z}_{1:k})}.
\end{align}

From \eqref{eq7}, the unnormalized weight $\upalpha_k(\bm{x}^i_{1:k})$ can be updated recursively as:
\begin{align}\notag
\upalpha_k(\bm{x}^i_{1:k})  
&\propto\frac{p(\bm{z}_k|{\bm{x}^i_k})p({\bm{x}^i_k}|{\bm{x}^i_{k-1}})p({\bm{x}^i_{0:k-1}}|\bm{z}_{1:k-1})}{q({\bm{x}^i_k}|{{\bm{x}^i_{0:k-1}},\ \bm{z}}_{1:k})q({\bm{x}^i_{0:k-1}}|\bm{z}_{1:k-1})}\notag\\
&={\upalpha_{k-1}(\bm{x}^i_{1:k-1})}\frac{p(\bm{z}_k|{\bm{x}^i_k})p({\bm{x}^i_k}|{\bm{x}^i_{k-1}})}{q({\bm{x}}^i_k|{{\bm{x}^i_{0:k-1}}, \bm{z}}_{1:k})}. \notag
\end{align}
\newline
Finally, let $w_k(\bm{x}^i_{1:k}) = \upalpha_k(\bm{x}^i_{1:k})/\sum_{j = 1}^{m} \upalpha_k(\bm{x}^j_{1:k})$ be the normalized weights, which we use to construct the PF estimator. 

To deal with the problem of degeneration, the situation where all but a few particles have zero importance weight, we can re-sample the particles. 
Ref. \cite{2005} discussed several re-sampling schemes in the PF and in this study we consider the most commonly used one, multinomial re-sampling scheme. That is, at time $k$, we first update $ m $ particles from last step by $\tilde{\bm{x}}^i_k \sim p_k(\cdot | \bm{x}^i_{k-1})$, where, to be consistent with the notation of \cite{2013}, we use $\tilde{\bm{x}}^i_k $ to denote the particles before resampling. Then, we draw $m$ paths from $\{\tilde{\bm{x}}^i_{1:k}, 1 \leq i \leq m$ \} with probability $w^i_k$, where we denote the records before resampling as: $\tilde{\bm{x}}^i_{1:k} = (\bm{x}^i_{1:k-1}, \tilde{\bm{x}}^i_k)$, and records after resampling as: $\bm{x}^i_{1:k}$.


In this study, we focus on the particular situation where the above re-sampling process is done at every step and we assign new weights to the re-sampled particles. Moreover, the total number of particles remains unchanged as $ m $.

\subsection{Notation}
For terminal time point $T$, the conditional density function in the hidden Markov model implies:
\begin{equation}\label{eq9}
p_T(\bm{x}_{1:T} | \bm{z}_{1:T}) \propto \prod\limits^T_{k=1} [p_k(\bm{x}_k|\bm{x}_{k-1})p_k(\bm{z}_k|\bm{x}_k)].
\end{equation}


The likelihood ratio in the importance sampling process is:
\begin{equation}\label{eq11}
L_T(\bm{x}_{1:T}) =\frac{p_T(\bm{x}_{1:T} | \bm{z}_{1:T})}{\prod\limits^T_{k=1}q_k(\bm{x}_k | x_{1:k-1})}.
\end{equation}

For computational convenience, we also define the following quantities:

\begin{equation}\label{eq13}
\bar{\upalpha}_k = \frac{1}{m}\sum\limits^m_{j=1}\upalpha_k(\tilde{\bm{x}}^j_{1:k}),
\end{equation}
where $\upalpha_k(\tilde{\bm{x}}^j_{1:k})$ is unnormalized weight of the $j^{th}$ particle path before resampling, and
\begin{equation}\label{eq15}
\begin{aligned}
H^i_k &= \frac{\bar{\upalpha}_1 \cdots \bar{\upalpha}_k}{\prod\limits^k_{l=1}\upalpha_l(\bm{x}^i_{1:l})}\\
\tilde{H}^i_k &= \frac{\bar{\upalpha}_1 \cdots \bar{\upalpha}_k}{\prod\limits^k_{l=1}\upalpha_l(\tilde{\bm{x}}^i_{1:l})}\\
\end{aligned}
\end{equation}
where $\upalpha_l(\bm{x}^i_{1:l})$ is unnormalized weight of the $i^{th}$ particle path after resampling.\\
\newline
Following the notation of \cite{2013}, we denote the ''ancestry origin'' of a particle by $A^i_k$ to keep track of it and it is defined as follows: $A^i_0 = i$ for all $1 \leq i \leq m$ by definition. 
If $\bm{x}^i_{1:k}$ and $\bm{x}^j_{1:l}$, $l > k$, share the same first state vector in time, (i.e. $\bm{x}^i_{1:k} = (\bm{x}^{i_1}_1, \cdots, \bm{x}^{i_k}_k$), $\bm{x}^j_{1:l} = (\bm{x}^{j_1}_1, \cdots, \bm{x}^{j_l}_l$), and $i_1 = j_1$) they have the same ancestral particle, which implies $A^j_l = A^i_k$. 

Finally, let
\begin{equation}\label{eq19}
\left\{
\begin{array}{lr}
\mathcal{F}_{2k-1} = \{\tilde{\bm{x}}^i_1: 1 \leq i \leq m\} \cup \\
\ \ \ \ \ \ \ \ \ \{(\bm{x}^i_l, \tilde{\bm{x}}^i_{l+1}, A^i_l): 1 \leq l < k, 1 \leq i \leq m\},\\
\mathcal{F}_{2k} = \mathcal{F}_{2k-1} \cup \{(\bm{x}^i_k, A^i_k): 1 \leq i \leq m\}.
\end{array}
\right.
\end{equation}
denote the history information generated by the $m$ particles at the $ k^{th} $ step. Our definition of such history is in line with the decomposition of variance in the following analysis, which is aimed at constructing a nice martingale structure when proving the asymptotic normality of $\hat{\bm{x}}_T - E(\bm{x}_T | \bm{z}_{1:T})$.

\section{Mathematical Analysis}
This section includes the main formal results that justify our approach. Due to space limitation here, complete proofs are given in the appendix.
\subsection{Theorem Statement}
We state our main theorem below and then give a proof in Sec.III.C. Let the function $\bm{u}_0$ and $ \bm{u}_{t} (\bm{x}_{1:t}) $ be as follows:\\
\begin{equation}\label{eq20}
 \left\{
\begin{array}{lr}
\bm{u}_0 = E[\bm{x}_{T}|\bm{z}_{1:T}] , \\
\bm{u}_k(\bm{x}_{1:k}) = E [\bm{x}_{T} L_{T} (\bm{x}_{1:k}) |\bm{x}_{1:k} ] ,\ \ \  1 \leq k \leq T
\end{array}
\right.
\end{equation}\
\begin{equation}\label{eq100}
g^*_k(\bm{x}_{1:k}) = \frac{E[\prod\limits^k_{l=1}\upalpha_l(\bm{x}_{1:l}) ]}{\prod\limits_{l=1}^k\upalpha_l(\bm{x}_{1:l})},
\end{equation}
and same as our previous work \cite{CISS}, we define
$ \bm{\Sigma} = \sum_{k=1}^{2T-1} \bm{\Sigma}_{k} $, where\\
\begin{equation}\label{eq21}
 \left\{
\begin{array}{lr}
\bm{\Sigma}_{2k-1} = E \{ ( \bm{u}_{k} (\bm{x}_{1:k})\bm{u}_{k} (\bm{x}_{1:k})^T - \\
 \ \ \ \ \ \ \ \ \ \ \ \ \bm{u}_{k-1} (\bm{x}_{1:k-1})\bm{u}_{k-1} (\bm{x}_{1:k-1})^T) g_{k-1}^* \}, \\
\bm{\Sigma}_{2k} = E \{ [\bm{u}_{k} (\bm{x}_{1:k})g_{k}^{*} - \bm{u}_{0}] {[\bm{u}_{k} (\bm{x}_{1:k})g_{k}^{*} - \bm{u}_{0}]}^{T} / g_{k}^{*} \}.\
\end{array}
\right.
\end{equation}
\newline
\textbf{Theorem:} 
Assume the HMP as \eqref{eq*}, for PF estimator $\hat{\bm{x}}_k$ obtained by resampling at each step, and $\text{det}(\bm{\Sigma}_k) < \infty$ for all $ k $. Then, 
$ \sqrt{m}(\hat{\bm{x}}_{T}-E[\bm{x}_{T}|\bm{z}_{1:T}]) \stackrel{\text{dist}}{\longrightarrow} N (\bm{0}, \bm{\Sigma})$, as $m \rightarrow \infty$.

\subsection{Estimator}
In the PF algorithm, the true estimator of $E[\bm{x}_T | \bm{z}_{1:T}]$ is:
\begin{equation}\label{eq22}
\hat{\bm{x}}_T = \sum\limits_{i=1}^m \tilde{\bm{x}}^i_T w_T(\tilde{\bm{x}}^i_{1:T}) = (m\bar{\upalpha}_T)^{-1}\sum\limits_{i=1}^m \tilde{\bm{x}}^i_T \upalpha_T(\tilde{\bm{x}}^i_{1:T}).
\end{equation}

To show asymptotically normality of $\hat{\bm{x}}_T - E[\bm{x}_T|\bm{z}_{1:T}]$, we need an estimator that has nice martingale properties to find its asymptotic variance.  Hence, we re-express \eqref{eq22} by \eqref{eq23} below based on the following rationale, and then prove \eqref{eq23} converges to \eqref{eq22} as $m$ gets large and has nice martingale structure. To facilitate that derivation, we first represent $\hat{\bm{x}}^*_T$ as
\begin{equation}\label{eq23}
\hat{\bm{x}}^*_T = m^{-1} \sum\limits_{i=1}^m L_T(\tilde{\bm{x}}^i_{1:T})\tilde{\bm{x}}^i_{T}H^i_{T-1}.
\end{equation}
 Furthermore, it is easier to derive the asymptotic variance of $ \hat{\bm{x}}^*_T - E[\bm{x}_T|\bm{z}_{1:T}]  $. \\
\newline
Note that  $\hat{\bm{x}}_T$ and $\hat{\bm{x}}^*_T$ have the same (normalized) limiting distribution. However, \eqref{eq23} can not be used in practice because it contains normalizing constants $L_T(\tilde{\bm{x}}^i_{1:T})$ which is often unknown.\\
\newline
Next, we provide the reasoning behind \eqref{eq23} as follows:
\begin{align}
E[\prod\limits_{t=1}^T \upalpha_t(\bm{x}_{1:t})]
& = \int\prod\limits_{t=1}^T \upalpha_t(\bm{x}_{1:t})q_t(\bm{x}_t|\bm{x}_{1:t-1}) d\nu(\bm{x}_{1:T})\notag\\
& = \int\prod\limits_{t=1}^T p_t(\bm{x}_t | \bm{x}_{t-1})p_t(\bm{z}_t|\bm{x}_t) d\nu(\bm{x}_{1:T}),\label{eq24}
\end{align}
where $\nu(\bm{x}_{1:T})$ is the measure defined on space of all records corresponding to probability density function $p_k(\cdot|\bm{x}_{k-1})$ (note that $p(\cdot|\bm{x}_{k-1}) = d P(\cdot|\bm{x}_{k-1}) / d \nu(\bm{x}_{1:T})$).\\

 Normalizing the right hand side of \eqref{eq9}, we know that:
 \[p_T(\bm{x}_{1:T} | \bm{z}_{1:T}) =  \frac{\prod\limits^T_{k=1} [p_k(\bm{x}_k|\bm{x}_{k-1})p_k(\bm{z}_k|\bm{x}_k)]}{ \int\prod\limits_{k=1}^T p_k(\bm{x}_k | \bm{x}_{k-1})p_t(\bm{z}_k|\bm{x}_k) d\nu(\bm{x}_{1:T})}. \] 

Combining with \eqref{eq24}, we have:
\[p_T(\bm{x}_{1:T} | \bm{z}_{1:T}) = \frac{\prod\limits^T_{k=1} [p_k(\bm{x}_k|\bm{x}_{k-1})p_k(\bm{z}_k|\bm{x}_k)]}{E[\prod\limits^T_{k=1}\upalpha_k(\bm{x}_{1:k}) ]}.\]
Then,
\begin{align}
L_T(\bm{x}_{1:T}) &= \frac{p_T(\bm{x}_{1:T}|\bm{z}_{1:T})}{\prod\limits^T_{k=1}q(\bm{x}_k | \bm{x}_{1:k})}
= \frac{\prod\limits^T_{k=1}\upalpha_k(\bm{x}_{1:T})}{E[\prod\limits^T_{k=1}\upalpha_k(\bm{x}_{1:k}) ]}  \notag \\
\ \ L_T(\tilde{\bm{x}}^i_{1:T})H^i_{T-1} &=\frac{\prod\limits^T_{k=1}\upalpha_k(\tilde{\bm{x}}^i_{1:k})\bar{\upalpha}_1 \cdots \bar{\upalpha}_{T-1}}{E[\prod\limits^T_{k=1}\upalpha_k(\bm{x}_{1:k}) ]\prod\limits^{T-1}_{k=1}\upalpha_k(\tilde{\bm{x}}^i_{1:k})} ,\notag\\
&= \frac{\bar{\upalpha}_1 \cdots \bar{\upalpha}_{T}}{E[\prod\limits^T_{k=1}\upalpha_k(\bm{x}_{1:k}) ]}\frac{\upalpha_T(\tilde{\bm{x}}^i_{1:T})}{\bar{\upalpha}_T}.\label{eq26}
\end{align}
From \eqref{eq23} and \eqref{eq26}, 

\begin{align}
&\upalpha_T(\tilde{\bm{x}}^i_{1:T}) = \frac{\bar{\upalpha}_TE[\prod\limits^T_{k=1}\upalpha_k(\bm{x}_{1:k}) ]}{\bar{\upalpha}_1 \cdots \bar{\upalpha}_T} L_T(\tilde{\bm{x}}^i_{1:T})H^i_{T-1},\label{eq27}\\
&\hat{\bm{x}}_T = (m\bar{\upalpha}_T)^{-1}\sum\limits_{i=1}^m \tilde{\bm{x}}^i_T \upalpha_T(\tilde{\bm{x}}^i_{1:T})\notag\\
&  = (m\bar{\upalpha}_T)^{-1}\sum\limits_{i=1}^m \tilde{\bm{x}}^i_T \frac{\bar{\upalpha}_TE[\prod\limits^T_{k=1}\upalpha_k(\bm{x}_{1:k}) ]}{\bar{\upalpha}_1 \cdots \bar{\upalpha}_T} L_T(\tilde{\bm{x}}^i_{1:T})H^i_{T-1}\notag\\
& = m^{-1} \frac{E[\prod\limits^T_{k=1}\upalpha_k(\bm{x}_{1:k}) ]}{\bar{\upalpha}_1 \cdots \bar{\upalpha}_T} \sum\limits_{i=1}^m \tilde{\bm{x}}^i_T L_T(\tilde{\bm{x}}^i_{1:T})H^i_{T-1}\notag\\
& =  \frac{E[\prod\limits^T_{k=1}\upalpha_k(\bm{x}_{1:k}) ]}{\bar{\upalpha}_1 \cdots \bar{\upalpha}_T} \hat{\bm{x}}^*_T.\label{eq28}
\end{align}

By lemma 2, which we will state and prove later, $\bar{\upalpha}_1 \cdots \bar{\upalpha}_T \stackrel{p}{\rightarrow} E[\prod^T_{k=1}\upalpha_k(\bm{x}_{1:k}) ]$ as $m \rightarrow \infty$, and we have that $\hat{\bm{x}}^*_T$ converges to the true PF estimator $\hat{\bm{x}}_T$ in probability as $m \rightarrow \infty$.

\subsection{Proof for Theorem}
Here, we provide a rigorous proof for the theorem of Sec.III.A. To begin with, we propose the following two lemmas.\\

\textbf{Lemma 1:}
Let $\bm{G}$ be a measurable vector function from history of state-space $\mathbb{R}^{t \times n}$ ($t$ is time and $n$ is dimension of state) to $\mathbb{R}^s$, where $s < \infty$.   For any $1 \leq k \leq T$, we define $g^*_k(\bm{x}_{1:k})$ as \eqref{eq100}. Then,\\
\newline
(i) if $\left\lVert E[\bm{G}(\bm{x}_{1:k}) / g_{k-1}^*(\bm{x}_{1:k-1})]\right\rVert_{\infty} < \infty$, where $\left\lVert \bm{y}\right\rVert_{\infty} = \max_i |\bm{y}_i|$ for any vector $\bm{y}$, as $m \rightarrow \infty$,
\[ m^{-1}\sum\limits_{i=1}^m \bm{G}(\tilde{\bm{x}}_{1:k}^i) \stackrel{p}{\longrightarrow} E[\bm{G}(\bm{x}_{1:k}) / g^*_{k-1}(\bm{x}_{1:k-1})]; \]
\newline
(ii) if $\left\lVert E[\bm{G}(\bm{x}_{1:k}) / g_{k}^*(\bm{x}_{1:k})]\right\rVert_{\infty} < \infty$, as $m \rightarrow \infty$, 
\[ m^{-1}\sum\limits_{i=1}^m \bm{G}(\bm{x}_{1:k}^i) \stackrel{p}{\longrightarrow} E[\bm{G}(\bm{x}_{1:k}) / g^*_{k}(\bm{x}_{1:k})]. \]

\textit{Proof:} Here we only give the lemma statement. All proof details are in the appendix.

\textbf{Lemma 2:}
If the same conditions in Lemma 1 are satisfied, then
\begin{align}
\frac{H^i_k}{g^*_k(\bm{x}^i_{1:k})} \stackrel{p}{\longrightarrow} 1\ \ \ as\ m \rightarrow \infty. \notag
\end{align}
Furthermore, if $\bm{G}$ is a vector function from $\mathbb{R}^{t\times n}$ ($t$ is time and $n$ is dimension of state) to $\mathbb{R}$, and $E[|\bm{G}(\bm{x}_{1:k})| / g^*_{k-1}(\bm{x}_{1:k-1})] < \infty$, then
\begin{align}
m^{-1}\sum\limits^m_{i=1}|\bm{G}(\tilde{\bm{x}}^i_{1:k})|\mathbb{1}_{\{|\bm{G}(\tilde{\bm{x}}^i_{1:k})| > \upepsilon \sqrt{m}\}} \stackrel{p}{\longrightarrow} 0\ \ \ \ as\ \ m\rightarrow \infty. \notag
\end{align}
\textit{Proof:} We only give the lemma statement here. All proof details are in the appendix.
\newline
\subsubsection{Conditional Distribution} 
First, let us consider the distribution of $ \sqrt{m}(\hat{\bm{x}}^*_{k}-E (\bm{x}_{T} |\bm{z}_{1:t} ))$ conditional on $\mathcal{F}_{k-1} $, and we will show that it is asymptotically normal as the number of particles $m$ gets large.


According to \eqref{eq15}, we have
\begin{equation}
m w_{t}^{i} = \frac{H_{t-1}^{i}}{\tilde{H}_{t}^{i}},\label{eq65}
\end{equation}
\begin{equation}
\sum_{i:A^{t}_{t}=j} \bm{u}_{t} (\bm{x}^{i}_{1:t}) H^{i}_{t}= \sum_{i:A^{t}_{t}=j} \#^{i}_{t}  \bm{u}_{t} (\tilde{\bm{x}}^{i}_{1:t}) \tilde{H}^{i}_{t}.\label{eq66}
\end{equation}
where $\#^{i}_{t}$ is the number of copies generated in the resampling process for particle path $\tilde{\bm{x}}^{i}_{1:t}$.

Combining \eqref{eq65} and \eqref{eq66}, we have:

\begin{align}
&\sum_{t=1}^{T} \sum_{i : A^{i}_{t-1} =j} [ \bm{u}_{t} (\tilde{\bm{x}}^{i}_{1:t}) - \bm{u}_{t-1} (\bm{x}^{i}_{1:t-1}) ] H^{i}_{t-1} \notag\\
&+ \sum_{t=1}^{T} \sum_{i : A^{i}_{t-1} =j} (\#^{i}_{t} - m w^{i}_{t}) \bm{u}_{t} (\tilde{\bm{x}}^{i}_{1:t})H^{i}_{t} \notag\\
&=\sum_{i : A_{t-1}^{i} =j} \bm{u}_{t} (\tilde{\bm{x}}^{i}_{1:T})H^{i}_{T-1} - \bm{u}_{0}. \label{eq67}
\end{align}

Plugging \eqref{eq23} into \eqref{eq67},
\begin{equation}\notag
m (\hat{\bm{x}}^*_{T} - E (\bm{x}_{T} |\bm{z}_{1:t} )) = \sum_{i=1}^{m} \bm{u}_{t} (\tilde{\bm{x}}^{i}_{1:T}) {H}^{i}_{T-1} - m\bm{u}_{0}.
\end{equation}

Then,
\begin{equation}\label{eq69}
m (\hat{\bm{x}}^*_{T} - E (\bm{x}_{T} |\bm{z}_{1:t} )) = \sum_{k=1}^{2T-1} \sum_{i=1}^{m} M^{i}_{k}.
\end{equation}
where
\begin{equation}\label{eq70}
\left\{
\begin{array}{lr}
M^{i}_{2t-1} = [ \bm{u}_{t} (\tilde{\bm{x}}^{i}_{1:t}) - \bm{u}_{t-1} (\bm{x}^{i}_{1:t-1}) ] H^{i}_{t-1},  \\
M^{i}_{2t} = \bm{u}_{t} (\bm{x}^{i}_{1:t})H^{i}_{t} - \sum_{j=1}^{m} w^{j}_{t}\bm{u}_{t} (\tilde{\bm{x}}^{j}_{1:t})\tilde{H}^{j}_{t}.
\end{array}
\right.
\end{equation}

Next, we prove that the above \eqref{eq70} is a martingale difference sequence. Firstly, according to the importance sampling, the conditional distribution of $ \tilde{\bm{x}}^i_{1:t} $ for $ 1 \leq i \leq m $ given $ \mathcal{F}_{2t-2} $ are independent with  $ \tilde{\bm{x}}^i_{1:t} $ having the density function $ q_t(.|x^i_{1:t-1}) $. Secondly, in the resampling process, the conditional distribution of $ \bm{x}^i_{1:t} $ for $ 1 \leq i \leq m $ given $ \mathcal{F}_{2t-1} $ are i.i.d that can take on the values $ \tilde{\bm{x}}^i_{1:t} $ with probability $ w^i_t $. 

From \eqref{eq70}, noticing that $\mathcal{F}_1$ contains information of $\tilde{\bm{x}}_1$, we have\\ \[ E (M^{j}_{2} | \mathcal{F}_{1} )=E \{ \bm{u}_{1} (\bm{x}_{1}^{i})H_{1}^{i} - \sum_{j=1}^{m} w_{1}^{j}\bm{u}_{t} (\tilde{\bm{x}}_{1}^{j})\tilde{H}^{j}_{1} | \tilde{\bm{x}}_1  \}= 0.\]\\
Also, we have\\
 \[ E (M_{3}^{j} | \mathcal{F}_{2} ) = E \{  [ \bm{u}_{2} (\tilde{\bm{x}}^{i}_{1:2}) - \bm{u}_{1} (\bm{x}^{i}_{1}) ] H^{i}_{1} | \mathcal{F}_{2}  \} = 0.\]\\
 Thus, carry out this process and finally we have that $ \{ {M}^{j}_{k}, \mathcal{F}_{k}, 1 \leq k \leq 2T-1 \} $ is a martingale difference sequence. Additionally, we have $ M_{k}^{1}, M_{k}^{2},..., M_{k}^{m} $ are independent conditioning on $ \mathcal{F}_{k-1} $.

Rearranging \eqref{eq69}, we have
\begin{equation}\notag
\sqrt{m}(\hat{\bm{x}}^*_{T} - E (\bm{x}_{T} |\bm{z}_{1:t} )) = \sum_{k=1}^{2T-1} (\sum_{i=1}^{m} \frac{M^{i}_{k}}{\sqrt{m}} ).
\end{equation}

Then, given any vector $ \uptheta $, we have
\begin{align}
&E[\uptheta^{T} M^{i}_{2t-1} {M^{i}_{2t-1}}^{T} \uptheta | \mathcal{F}_{2t-2} ] \notag \\
&= \uptheta^{T} \{ E[\bm{u}_{t} (\tilde{\bm{x}}^{i}_{1:t}) {\bm{u}^{T}_{t} (\tilde{\bm{x}}^{i}_{1:t})}|\bm{x}^{i}_{1:t-1}  ] \notag\\ 
&\ \ \ \ - \bm{u}_{t-1} (\bm{x}^{i}_{1:t-1}) {\bm{u}^{T}_{t-1} (\bm{x}^{i}_{1:t-1})}  \} \uptheta {H^{i}_{t-1}}^{2} < \infty, \label{eq72}
\end{align}
\begin{align}
&E[\uptheta^{T} M^{i}_{2t} {M^{i}_{2t}}^{T} \uptheta | \mathcal{F}_{2t-1} ] = \uptheta^{T}  \Big\{ \sum_{j=1}^{m} w^{j}_{t} \bm{u}_{t} (\tilde{\bm{x}}^{i}_{1:t}) {\bm{u}^{T}_{t} (\tilde{\bm{x}}^{j}_{1:t})} { {\tilde{H}}^{j}_{t}}\notag \\
&\ \ \ \ \ \ \ \ \ \ \ \ \ \ \ \   -{  [ \sum_{j=1}^{m} w_{t}^{j} \bm{u}_{t} (\tilde{\bm{x}}^{i}_{1:t}) {\bm{u}_{t} (\tilde{\bm{x}}^{j}_{1:t})} {\tilde{H}}^{j}_{t}  ] }^{2}  \Big\} \uptheta < \infty. \label{eq73}
\end{align}

By \eqref{eq72} and \eqref{eq73}, for any $\upepsilon > 0 $, applying lemma 1 and 2, we have
\begin{equation}\label{eq74}
\sum_{i=1}^{m} E \left [ \left. \frac{1}{m}  M^{i}_{k} {M^{i}_{k}}^{T} \right | {\mathcal{F}}_{k-1} \right ] \stackrel{p}{\rightarrow}  \bm{\Sigma}_{k}, 
\end{equation}

\begin{equation}\label{eq75}
\sum_{j=1}^{m} E \left [ \left. \frac{  {\uptheta}^{T}{M^{i}_{k}}{M^{i}_{k}}^{T}\uptheta   }{m}  \mathbb{1}_{\left \{ \left |{\uptheta}^{T} {M^{i}_{k}}/\sqrt{m} \right | > \upepsilon \right \} } \right | {\mathcal{F}}_{k-1} \right ] \stackrel{p}{\rightarrow} 0
\end{equation}
\\
as $ m \rightarrow \infty$\\
Therefore, by multivariate Lindeberg's central limit theorem \cite{LIND}, the conditional distribution converges to normal distribution:
\begin{align}
 \sqrt{m}(\hat{\bm{x}}^*_{T}-E (\bm{x}_{T} |\bm{z}_{1:T} )) = \sum_{i=1}^{m} \frac{M^{i}_{k}}{\sqrt{m}} \notag \\
\stackrel{\mathcal{F}-\mathcal{D}}{\longrightarrow} N (\bm{0}, \bm{\Sigma}  )\ as\  m \rightarrow \infty. \label{eq76}
\end{align}
Where $\stackrel{\mathcal{F}-\mathcal{D}}{\longrightarrow}$ means that $\sqrt{m}(\hat{\bm{x}}^*_{T}-E (\bm{x}_{T} |\bm{z}_{1:T} ))$ conditional on $\mathcal{F}$ converge in distribution to $N (0, \bm{\Sigma})$.
Although of $P(\hat{\bm{x}}^*_{T}-E (\bm{x}_{T} |\bm{z}_{1:T})| \mathcal{F}_{T-1})$ is a function of history records, as the number of particles increase, the distribution of $\hat{\bm{x}}^*_{T}-E (\bm{x}_{T}|\bm{z}_{1:T})$ conditional on $ \mathcal{F}_{T-1}$   becomes stable. Therefore, the variance of this distribution is a constant.

\subsubsection{Unconditional Distribution} 

In this part, we want to show that $ \sqrt{m}(\hat{\bm{x}}^*_{T}-E (\bm{x}_{T} |\bm{z}_{1:T} )) $ is asymptotically normal.
In particular, our goal is to show unconditional asymptotic distribution of $ \sqrt{m}(\hat{\bm{x}}^*_{T}-E (\bm{x}_{T} |\bm{z}_{1:T} )) $ converges to the same characteristic function as that of normal distribution.
\begin{equation}\label{eq78}
E \left [  e^{iu^{T} \sqrt{m}(\hat{\bm{x}}^*_{k}-E (\bm{x}_{k} |\bm{z}_{1:k} )) }\right ]\stackrel{p}{\rightarrow} e^{ -\frac{1}{2} u^{T} \bm{\Sigma} u  }. 
\end{equation}

Specifically, we want to prove that:
\begin{align}\label{eq79}
&E \left [  e^{iu^{T} \sqrt{m}(\hat{\bm{x}}^*_{k}-E (\bm{x}_{k} |\bm{z}_{1:k} )) }\right ]\notag\\
&= E \left [  e^{iu^{T} \sum_{k=1}^{2T-2} ( \sum_{j=1}^{m} M^{j}_{k} / \sqrt{m} ) } E \left( \left. e^{ \sum_{j=1}^{m}  M^{j}_{2T-1} / \sqrt{m}}  \right | \mathcal{F}_{2T-2} \right) \right ] \notag\\
&= e^{ -\frac{1}{2} u^{T} \bm{\Sigma}_{2T-1} u} E \left( e^{iu^{T} \sum_{k=1}^{2T-2} ( \sum_{j=1}^{m} M^{j}_{k} / \sqrt{m} ) }  \right) + o(1). 
\end{align}

The proof details are shown in the appendix. In supporing matrial, we showed that \eqref{eq78} holds. Therefore, the unconditional distribution of $ \sqrt{m}(\hat{\bm{x}}^*_{T}-E (\bm{x}_{T} |\bm{z}_{1:T} )) $ is asymptotically normal with mean zero and covariance matrix $ \bm{\Sigma}$. Since we have shown that, $\hat{\bm{x}}^*_{T} \stackrel{p}{\rightarrow}  \hat{\bm{x}}_T $, we can conclude that $ \sqrt{m}(\hat{\bm{x}}_{T}-\bm{x}_{T}) \stackrel{\text{dist}}{\longrightarrow} N (\bm{0}, \bm{\Sigma})$.

\section{Numerical Study}
In this part, let us test the asymptotic normality of $ \sqrt{m}(\hat{\bm{x}}_{T}-E (\bm{x}_{T} |\bm{z}_{1:T} )) $ by studying the following two examples. One example has linear state and measurement equations but non-Gaussian noise terms. The other example is the classic multivariate stochastic volatility model in finance. It is a nonlinear and Gaussian system. The selection of these two example considers the simplicity of interpretation and wide-spread use of similiar system.

\subsection{Linear and Non-Gaussian}
Consider the following linear DSSM with non-Gaussian measurement error,
\[ \bm{x}_{k+1}=\left[\begin{matrix}0.5&0.5&0\\0&0.5&0.5\\0.5&0&0.5\\\end{matrix}\right]\bm{x}_k+\bm{w}_k, \]
\[ \textbf{z}_k=\left[\begin{matrix}\begin{matrix}0.5\\0.5\\\end{matrix}&\begin{matrix}0.5\\0.5\\\end{matrix}&\begin{matrix}0.5\\0.5\\\end{matrix}\\\end{matrix}\right]\bm{x}_k+\bm{v}_k.
\]
where $ \bm{w}_{k} $ and $ \bm{v}_{k} $ are independent noise terms with each component following a $ U_{[-1,1]} $ distribution. We approximate the conditional mean by PF using $ 10^6 $ particles. Meanwhile, we generate the PF estimators by the algorithm using $ 10^3 $ particles.

We run the simulation for 500 times with terminal $T = 25$. The histograms of each component of $ \hat{\bm{x}}_{k}-E (\bm{x}_{25} |\bm{z}_{1:25} ) $ are follows:

\begin{figure}[ht]
	\centering
	\includegraphics[width=.85\linewidth]{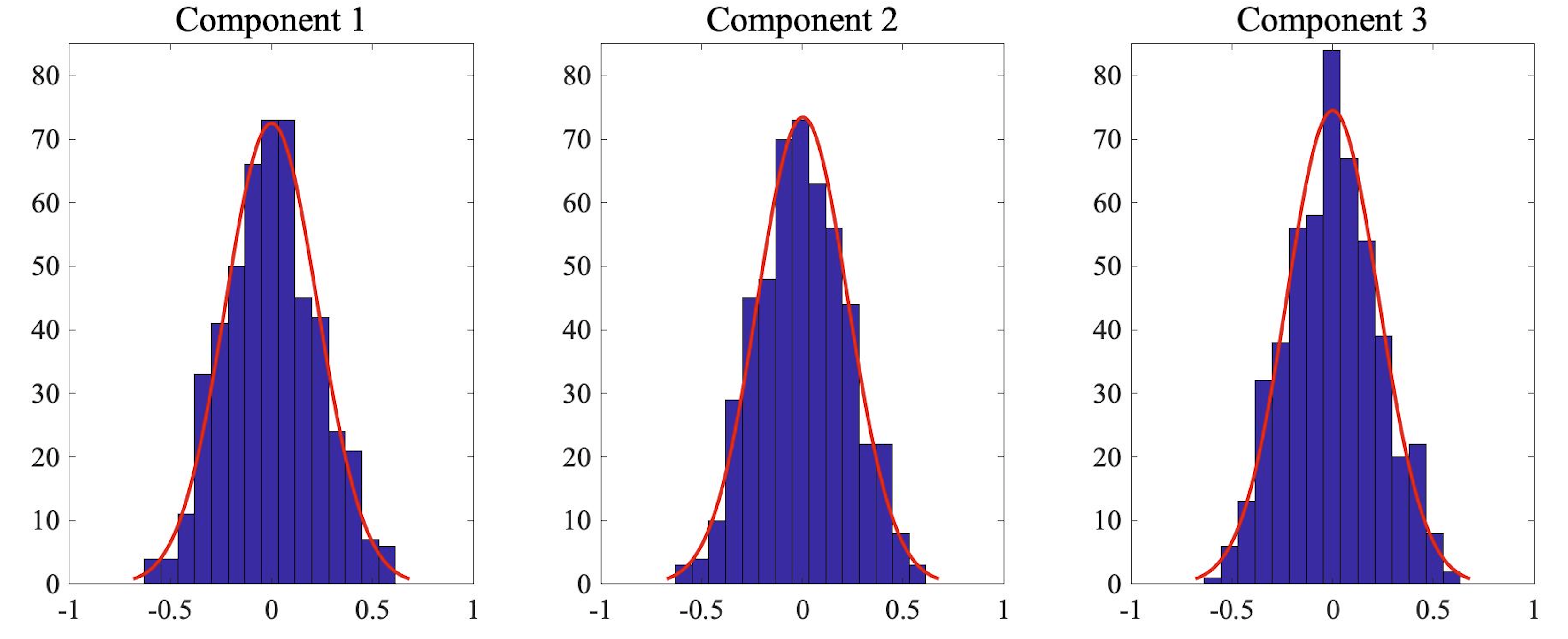}
	\caption{Histogram of Each Component in $ \tilde{\bm{x}}_{25}-E (\bm{x}_{25} |\bm{z}_{1:25} ) $}
	\label{fig:e1}
\end{figure}

Using the Jarque-Bera test of normality \cite{jbest}, the $p$-values for each component in the error term are respectively 0.2795, 0.2438 and 0.2138. Then, we do not reject the null hypothesis of normality for all three components at the significance level of 0.05 (no adjustment for multiple comparisons here).

This result is in line with our theorem that $ \hat{\bm{x}}_{k}-E (\bm{x}_{k} |\bm{z}_{1:k} ) $ follows an asymptotic normal distribution.

\subsection{Nonlinear and Gaussian}
Next, let us consider a slightly more complex model: Multivariate Stochastic Volatility model. As stated in \cite{2006}, it is a classical approach to model the underlying volatility of financial derivatives using observable variables.
Let $ \bm{x}_{t} $ denote the volatility vector, and $ \bm{z}_{t} $ be the observation vector. The system can be expressed as:
\[ \bm{x}_{k+1}=\upmu+\Phi\left(\bm{x}_k-\upmu\right)+\bm{w}_k,
\]
\[ \bm{z}_k=\left[\begin{matrix}\text{exp}(\frac{\bm{x}_{k1}}{2})&\cdots&0\\\vdots&\ddots&\vdots\\0&\cdots&\text{exp}(\frac{\bm{x}_{kp}}{2})\\\end{matrix}\right]\bm{v}_k.
\]
where $ \upmu $ denote the mean of state vector and $ \Phi $ denote a matrix with each element being constant. $ \bm{w}_t $ and $ \bm{v}_t $ denote the multivariate normal noise terms.

This this example, consider the case when $\mu = [0,0,0]^T, \Phi = 0.5, p = 3, w_k\ and\ v_k$ being standard normal. Similarly, conditional mean is approximated by PF using $ 10^6 $ particles and we generate the PF estimators by the algorithm using $ 500 $ particles.

Run the simulation for 500 times and the histograms of each component of $ \hat{\bm{x}}_{25}-E (\bm{x}_{25} |\bm{z}_{1:25} ) $ is shown as follows:

\begin{figure}[ht]
	\centering
	\includegraphics[width=.85\linewidth]{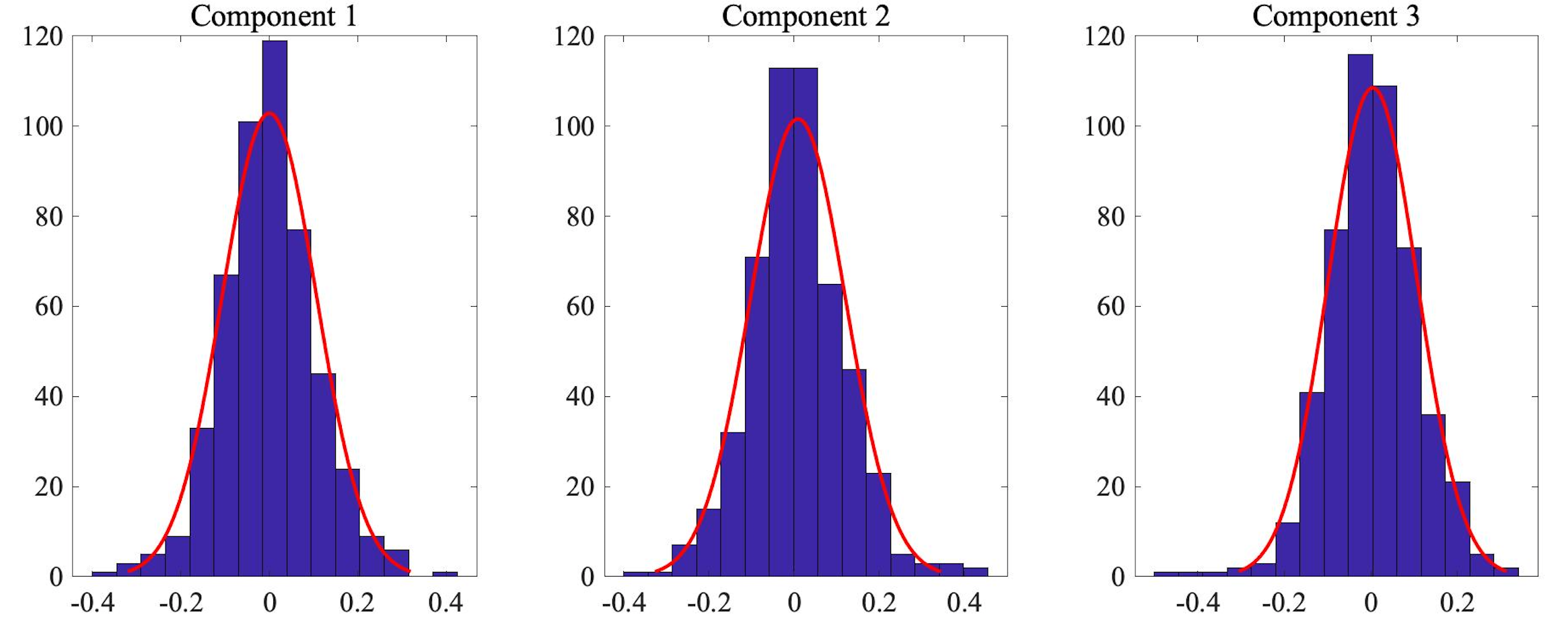}
	\caption{Histogram of Three Components in $ \hat{\bm{x}}_{25}-E (\bm{x}_{25} |\bm{z}_{1:25} ) $}
	\label{fig:e2}
\end{figure}

This time, the $p$-value for each component in the error term are respectively 0.0584, 0.1799 and 0.8063. Thus, we do not reject the null hypothesis of normality for all three components at the significant level of 0.05 (again, no adjustment for multiple comparisons here).

\section{CONCLUSIONS AND DISCUSSION}

From the above analysis and numerical results, we come to the conclusion that $ \hat{\bm{x}}_T - E[\bm{x}_T|\bm{z}_{1:T}] $ is asymptotically normal as the number of particles $ m $ gets sufficiently large. For further work, we will consider a computable approximation for the covariance matrix in the asymptotic distribution, which we discuss in  another work \cite{CISS}. Moreover, as is stated in the framework \eqref{important}, since the second part of the error decomposition is not necessarily normal, we will focus more on providing a reasonable bound for it.

\section*{Appendix}
\subsection{Lemma 1:}
Let $\bm{G}$ be a measurable vector function from history of state-space $\mathbb{R}^{t \times n}$ ($t$ is time and $n$ is dimension of state) to $\mathbb{R}^s$, where $s < \infty$.   For any $1 \leq k \leq T$, we define $g^*_k(\bm{x}_{1:k})$ as following:
\[g^*_k(\bm{x}_{1:k}) = \frac{E[\prod\limits^k_{l=1}\upalpha_l(\bm{x}_{1:l}) ]}{\prod\limits_{l=1}^k\upalpha_l(\bm{x}_{1:l})},\]
where $\upalpha(\bm{x}_{1:l})$ is unnormalized weight defined in equation \eqref{eq7}. Then,\\
\newline
(i) if $\left\lVert E[\bm{G}(\bm{x}_{1:k}) / g_{k-1}^*(\bm{x}_{1:k-1})]\right\rVert_{\infty} < \infty$, where $\left\lVert \bm{y}\right\rVert_{\infty} = max_i |\bm{y}_i|$ for any vector $\bm{y}$, as $m \rightarrow \infty$,
\[ m^{-1}\sum\limits_{i=1}^m \bm{G}(\tilde{\bm{x}}_{1:k}^i) \stackrel{p}{\longrightarrow} E[\bm{G}(\bm{x}_{1:k}) / g^*_{k-1}(\bm{x}_{1:k-1})]; \]
\newline
(ii) if $\left\lVert E[\bm{G}(\bm{x}_{1:k}) / g_{k}^*(\bm{x}_{1:k})]\right\rVert_{\infty} < \infty$, as $m \rightarrow \infty$, 
\[ m^{-1}\sum\limits_{i=1}^m \bm{G}(\bm{x}_{1:k}^i) \stackrel{p}{\longrightarrow} E[\bm{G}(\bm{x}_{1:k}) / g^*_{k}(\bm{x}_{1:k})]. \]

\textit{Proof:}
This lemma can be proved by induction: first, we show that if (ii) holds for $k-1$, then (i) holds for $k$. Then, we prove that if (i) holds for $k$, (ii) holds for $k$ using the same method. 

We declare some notation first for computational convenience: for any two real value function $f(x)$ and $g(x)$, $f^+(x) = \max(f(x), 0)$ for all $x$, $f^-(x) = - \min(f(x), 0)$ for all $x$, and $f(x) \wedge g(x) = \min(f(x), g(x))$ for all $x$. For any two function $ \upphi $ and $ \upgamma $ from $\mathbb{R}^n$ to $\mathbb{R}^s$, define:
\begin{align}
&\upphi \wedge \upgamma = (\upphi_1 \wedge \upgamma_1, \cdots, \upphi_s \wedge \upgamma_s),\label{eq34}\\
&\upphi^+ = (\upphi_1^+, \cdots, \upphi_s^+).\label{eq35}
\end{align}
$\bm{G}(\bm{r}) = (\upgamma_1(\bm{r}), \dots, \upgamma_s(\bm{r}))$ where for all $1 \leq i \leq s$, $\upgamma_i(\bm{r})$ is a real-valued function defined on $\mathbb{R}^n$. Define $\bm{G}^+(\bm{r}) = (\upgamma_1^+(\bm{r}), \dots, \upgamma_s^+(\bm{r}))$, $\bm{G}^-(\bm{r}) = (\upgamma_1^-(\bm{r}), \dots, \upgamma_s^-(\bm{r}))$, then we have $\bm{G}(\bm{r}) = \bm{G}^+(\bm{r}) - \bm{G}^-(\bm{r})$. So, without loss of generality, we can assume that $\bm{G}_i(\bm{r}) \geq 0$ for all $i$.

First, check that (i) holds for $k = 1$. For this, notice that when $k = 1$ , $g^*_{k-1} \equiv 1$. Then, (i) can be expressed as:
\begin{align}
&m^{-1}\sum\limits_{i=1}^m \bm{G}(\tilde{\bm{x}}^i_{1}) \stackrel{p}{\longrightarrow} E[\bm{G}(\bm{x}_{1})].\notag 
\end{align}

Notice that $\upgamma_j(\bm{x}_1)$ has finite variance since $\|E[\bm{G}(\bm{x}_1)]\|_{\infty}  < \infty$ (by (i)). By the weak law of large numbers,
$m^{-1}\sum_{i=1}^m\upgamma_j(\tilde{\bm{x}}^i_{1})\stackrel{p}{\rightarrow} E[\upgamma_j(\bm{x}_{1})]$  for all $ 1 \leq j \leq s$. 

Thus, 
\begin{align}
&\left[\frac{1}{m}\sum\limits_{i=1}^m\upgamma_1(\tilde{\bm{x}}^i_{1}), \cdots, \frac{1}{m}\sum\limits_{i=1}^m\upgamma_s(\tilde{\bm{x}}^i_{1})\right] \notag\\
&\stackrel{p}{\longrightarrow} (E[\upgamma_1(\bm{x}_{1})], \cdots, E[\upgamma_s(\bm{x}_{1})]).\notag
\end{align}

Equivalently, we have
\begin{align}
&m^{-1}\sum\limits_{i=1}^m \bm{G}(\tilde{\bm{x}}^i_{1}) \stackrel{p}{\longrightarrow} E[\bm{G}(\bm{x}_{1})]. \notag
\end{align}

Next, we want to show that if (ii) holds for $k-1$, then (i) holds for $k$. This can be proved by a contradictory argument.

Denote $\upmu_k = E[\bm{G}(\bm{x}_{1:k}) / g^*_{k-1}(\bm{x}_{1:k-1})]$. And, in contrast to (i), there exists
$  m_1 < m_2 < \cdots$, $m_l, \cdots \rightarrow \infty$, $\upepsilon > 0, \updelta > 0$ such that for all $\ m   \in \{m_1, \cdots, m_l, \cdots \}$,
\begin{align}
P[\| m^{-1}\sum\limits_{i=1}^m \bm{G}(\tilde{\bm{x}}_{1:k}^i)  - \upmu_k\|_2> \upepsilon] > \updelta,\notag
\end{align}
 In fact, we can find $\upepsilon > 0$ such that for all $\ m   \in \{m_1, \cdots, m_l, \cdots \}$,
\begin{align}
P[\| m^{-1}\sum\limits_{i=1}^m \bm{G}(\tilde{\bm{x}}_{1:k}^i)  - \upmu_k\|_2 > \upepsilon] > \upepsilon(3 + \zeta_k). \label{eq33}
\end{align}
where $\zeta_k = E[\| \bm{G}(\tilde{\bm{x}}_{1:k})\|_2  / g^*_{k-1}(\bm{x}_{1:k-1})] $. The reason behind above equation is: as $\upepsilon \rightarrow 0^+$, the left hand side of \eqref{eq33} increases to 1 and the right hand side of \eqref{eq33} decreases to 0. Thus,  we can always find some $\upepsilon > 0$ to satisfy \eqref{eq33}.\\
\newline
Now, let us decompose $\bm{G}(\tilde{\bm{x}}^i_{1:k})$ into following three parts, which are easier to be computed and bounded:

By \eqref{eq34} and \eqref{eq35}, $\upphi = (\upphi\wedge \upgamma) + (\upphi - \upgamma)^+$. Thus, we can write $\bm{G}(\tilde{\bm{x}}^i_{1:t})$ as:
\begin{align}
\bm{G}(\tilde{\bm{x}}^i_{1:k}) =\textbf{U}_{k}^i + \textbf{V}_{k}^i + \textbf{S}_{k}^i. \label{eq36}
\end{align}
where
\begin{align} \label{eqK}
\bm{U}_{k}^i & = \bm{G}(\tilde{\bm{x}}^i_{1:k}) \wedge \bm{K} - E[\bm{G}(\tilde{\bm{x}}^i_{1:k}) \wedge \bm{K} | \mathcal{F}_{2k - 2}],\notag\\
\bm{V}_{k}^i & = E[\bm{G}(\tilde{\bm{x}}^i_{1:k}) \wedge \bm{K} | \mathcal{F}_{2k - 2}],\notag\\
\bm{S}_{k}^i & = [ \bm{G}(\tilde{\bm{x}}^i_{1:k}) - \bm{K}]^+,\notag\\
\bm{K} &= (\frac{\upepsilon^3 m}{\sqrt{s}}, \cdots, \frac{\upepsilon^3 m}{\sqrt{s}}) \in \mathbb{R}^s.
\end{align}

Note the following facts:
\begin{align}
&E[\bm{U}^i_{k}] = E[\bm{G}(\tilde{\bm{x}}^i_{1:k}) \wedge \bm{K}] - E[E[\bm{G}(\tilde{\bm{x}}^i_{1:k}) \wedge \bm{K} | \mathcal{F}_{2k - 2}]] = 0,\notag\\
&\mathrm{Cov}[\bm{U}^i_{k}, \bm{U}^j_{k}| \mathcal{F}_{2k - 2} ] = 0\ \  \forall i\neq j.\notag
\end{align}

Then,
\begin{align}
&P\{\| m^{-1} \sum\limits_{i=1}^m \bm{U}_{k}^i \|_2 \geq \upepsilon | \mathcal{F}_{2k-2}\}\notag\\
=& P\{\| m^{-1} \sum\limits_{i=1}^m \bm{U}_{k}^i \|^2_2 \geq \upepsilon^2 | \mathcal{F}_{2k-2}\}\notag\\
\leq &\frac{1}{m^2\upepsilon^2}\mathrm{Trace}(\mathrm{Cov}[\sum\limits_{i=1}^m \bm{U}^i_{k} | \mathcal{F}_{2k-2}])\notag\\
=& \frac{1}{m^2\upepsilon^2} \sum\limits_{i=1}^m\mathrm{Trace}(\mathrm{Cov}[ \bm{U}^i_{k} | \mathcal{F}_{2k-2}])\notag\\
=& \frac{1}{m^2\upepsilon^2} \sum\limits_{i=1}^m E[\| \bm{U}^i_{k}\|^2_2 | \mathcal{F}_{2k-2}].\label{eq43}
\end{align}

Since $\bm{U}^i_{k}$ is a projection of $\bm{G}(\tilde{\bm{x}}^i_{1:k})\wedge \bm{K}$ to the orthogonal subspace of $\mathcal{F}_{2k-2}$ (it means: $E[\bm{U}^i_k | \mathcal{F}_{2k-2}] = 0$), we have: 
\begin{align}\label{eq44}
\| \bm{U}^i_{k}\|^2_2 \leq \| \bm{G}(\tilde{\bm{x}}^i_{1:k})\wedge \bm{K} \|^2_2  \leq \| \bm{G}(\tilde{\bm{x}}^i_{1:k})\|_2 \cdot \| \bm{K} \|_2.
\end{align}

By \eqref{eq43}, \eqref{eq44}, and \eqref{eqK},
\begin{align}
&P\{\| m^{-1} \sum\limits_{i=1}^m \bm{U}_{k}^i \|_2 \geq \upepsilon \big |  \mathcal{F}_{2k-2}\}\notag\\
&\leq \frac{1}{m^2\upepsilon^2} \sum\limits_{i=1}^m E[\| \bm{U}^i_{k}\|^2_2 \huge{|} \mathcal{F}_{2k-2}]\notag\\
\leq & \frac{1}{m^2\upepsilon^2}\sum\limits_{i=1}^m \|\bm{K}\|_2 E[\|\bm{G}(\tilde{\bm{x}}^i_{1:k}) \|_2 \big | \mathcal{F}_{2k-2}]\notag\\
=&\frac{\upepsilon}{m}\sum\limits_{i=1}^m E[\| \bm{G}(\tilde{\bm{x}}^i_{1:k})\|_2 \big | \mathcal{F}_{2k-2}].\label{eq45}
\end{align}

Applying (ii) to $\bm{G}^*(\bm{x}_{1:k-1}) = E[\bm{G}(\tilde{\bm{x}}_{1:k}) | \mathcal{F}_{2k-2}]$, we have 
\begin{align} 
&E[E[\| \bm{G}(\tilde{\bm{x}}_{1:k})\|_2 | \mathcal{F}_{2k-2}] / g^*_{k-1}(\bm{x}_{1:k-1})] \notag \\
&=  E[E[\| \bm{G}(\tilde{\bm{x}}_{1:k})\|_2  / g^*_{k-1}(\bm{x}_{1:k-1})| \mathcal{F}_{2k-2}]] \notag\\
&=  E[\| \bm{G}(\tilde{\bm{x}}_{1:k})\|_2  / g^*_{k-1}(\bm{x}_{1:k-1})]= \zeta_k, \notag
\end{align}
and
\begin{align}
&\frac{1}{m}\sum\limits_{i=1}^m E[\| \bm{G}(\tilde{\bm{x}}^i_{1:k})\|_2 \big | \mathcal{F}_{2k-2}] \notag\\
& \stackrel{p}{\longrightarrow} E\big[E[\| \bm{G}(\tilde{\bm{x}}_{1:k})\|_2 \big | \mathcal{F}_{2k-2}] / g^*_{k-1}(\bm{x}_{1:k-1})\big] = \zeta_k. \label{eq47}
\end{align}

From \eqref{eq45} and \eqref{eq47}, it follows that for m sufficiently large,
\begin{equation}
P\bigg(\|\frac{1}{m}\sum\limits_{i=1}^m \bm{U}^i_{k}\|_2 > \upepsilon \bigg) \leq \upepsilon(1 + \zeta_k). \label{eq48}
\end{equation}

Using the same trick by applying (ii) to $\bm{G}^*(\bm{x}_{1:k-1}) = E[\bm{G}(\tilde{\bm{x}}_{1:k}) | \mathcal{F}_{2k-2}]$,we can prove:
\begin{equation}
\frac{1}{m}\sum\limits_{i=1}^m \bm{V}^i_{k} \longrightarrow \frac{1}{m}\sum\limits_{i=1}^m E[\bm{G}(\tilde{\bm{x}}^i_{1:k}) | \mathcal{F}_{2k-2}]\stackrel{p}{\longrightarrow} \upmu_k. \label{eq49}
\end{equation}

Applying (ii) to \[\bm{G}(\bm{x}_{1:k-1}) = E[\max\limits_j \vert \bm{G}(\bm{x}_{1:k})\vert_j \mathbb{1}_{\{ \max\limits_j\vert G(\bm{x}_{1:k})\vert_j  > n_l\} }\Big |\mathcal{F}_{2k-2}].\]

Then, $ as\ n_l \rightarrow \infty  $
\begin{align}
&\frac{1}{m}\sum\limits_{i=1}^mE[\max\limits_j \vert \bm{G}(\tilde{\bm{x}}^i_{1:k})\vert_j \mathbb{1}_{\{ \max\limits_j\vert \bm{G}(\tilde{\bm{x}}^i_{1:k})\vert_j  > n_l\} }\Big|\mathcal{F}_{2k-2}] \stackrel{p}{\longrightarrow} \notag\\
& E[\max\limits_j \vert \bm{G}(\bm{x}_{1:k})\vert_j \mathbb{1}_{\{ \max\limits_j\vert \bm{G}(\bm{x}_{1:k})\vert_j  > n_l\}} / g^*_{k-1}(\bm{x}_{1:k-1})]\notag\\
& \quad \quad \quad \quad \quad \quad \quad \quad \quad \quad \quad \quad \quad \quad \quad \quad \ \ \ \  \longrightarrow 0.\ \notag
\end{align}

In particular, we can choose $n_l$ such that
\begin{align}
P\{E[\max\limits_j \vert \bm{G}(\bm{x}_{1:l})\vert_j \mathbb{1}_{\{ \max\limits_j\vert \bm{G}(\bm{x}_{1:k})  > n_l\}} \Big| 
\mathcal{F}_{2k-2}] >2l^{-1}\} \leq \frac{1}{l}. \notag
\end{align}

Then,
\begin{align}
&\sum\limits_{i=1}^mP\{\bm{S}^i_{k} \neq 0 | \mathcal{F}_{2k-2}\} \notag\\
&= \sum\limits^i_{m} P\{\max\limits_j \vert \bm{G}(\tilde{\bm{x}}^i_{1:k}) \vert_j >  \frac{\upepsilon^3 m}{\sqrt{s}} \Big| \mathcal{F}_{2k-2}\}\notag\\
& \leq \frac{\sqrt{s}}{\upepsilon^3 m}\sum\limits_{i=1}^m E[\max\limits_j\vert \bm{G}(\tilde{\bm{x}}^i_{1:k})\vert_j \mathbb{1}_{\{ \max\limits_j \vert \bm{G}(\tilde{\bm{x}}^i_{1:k}) \vert_j >  \frac{\upepsilon^3 m}{\sqrt{s}}\}} \Big| \mathcal{F}_{2k-2}]. \notag
\end{align}

To assure this result, we require $\sqrt{s} n_l < m \upepsilon^3$, but this can be achieved by choosing $m_l$ large enough. Thus, with probability at least $1 - s/l$,
\begin{align}
\sum\limits_{i=1}^m P\{\bm{S}^i_{k} \neq 0 | \mathcal{F}_{2k-2}\} \leq \frac{2\sqrt{s}}{\upepsilon^3 l} \rightarrow 0\ , as\ l \rightarrow \infty.  \label{eq53}
\end{align}

Combining \eqref{eq48}, \eqref{eq49} and \eqref{eq53}, we can conclude that:
\begin{align}
&P\{\|m^{-1}\sum\limits_{i=1}^m \bm{G}(\tilde{\bm{x}}^i_{1:k}) - \upmu_k \|_2 > 3\upepsilon\} \notag\\
\leq & P\{\| m^{-1}\sum\limits_i \bm{U}^i_{k} \|_2 + \| m^{-1}\sum\limits_i \bm{V}^i_{k} - \upmu_k\|_2 + \| m^{-1}\sum\limits_i \bm{S}^i_{k} \|_2 > 3\upepsilon\} \notag\\
\leq & P\{\| m^{-1}\sum\limits_i \bm{U}^i_{k} \|_2 > \upepsilon \} + P\{\| m^{-1}\sum\limits_i \bm{V}^i_{k} - \upmu_k \|_2 > \upepsilon \} \notag \\
&\ \ \ + P\{ \| m^{-1}\sum\limits_i \bm{S}^i_{k} \|_2 > \upepsilon\} \notag\\
\leq &\upepsilon(1 + \zeta_k + 1 + 1) \leq \upepsilon(3 + \zeta_k). \notag
\end{align}

Contradiction! 
Therefore, If (ii) for $k - 1$, (i) for $k$. Analogously, we can prove that if (i) holds for $k$, (ii) holds for $k$.

\subsection{Lemma 2:}
If the same conditions in Lemma 1 are satisfied, then
\begin{align}
\frac{H^i_k}{g^*_k(\bm{x}^i_{1:k})} \stackrel{p}{\longrightarrow} 1\ \ \ as\ m \rightarrow \infty. \notag
\end{align}
$H^i_k$ in above equation is defined by \eqref{eq15}. Furthermore, if $\bm{G}$ is a vector function from $\mathbb{R}^{t\times n}$ ($t$ is time and $n$ is dimension of state) to $\mathbb{R}$, and $E[|\bm{G}(\bm{x}_{1:k})| / g^*_{k-1}(\bm{x}_{1:k-1})] < \infty$, then
\begin{align}
m^{-1}\sum\limits^m_{i=1}|\bm{G}(\tilde{\bm{x}}^i_{1:k})|\mathbb{1}_{\{|\bm{G}(\tilde{\bm{x}}^i_{1:k})| > \upepsilon \sqrt{m}\}} \stackrel{p}{\longrightarrow} 0\ \ \ \ as\ \ m\rightarrow \infty. \notag
\end{align}

\textit{Proof:} Considering the special case when $\bm{G} = \upalpha_k$, by Lemma 1 (i), we have
\begin{align}
&\bar{\upalpha}_k \stackrel{p}{\longrightarrow} E[\upalpha_k(\bm{x}_{1:k}) / g^*_{k-1}(\bm{x}_{1:k-1})], \notag\\
&\upalpha_k(\bm{x}_{1:k}) / g^*_{k-1}(\bm{x}_{1:k-1}) = \upalpha(\bm{x}_{1:k})  \frac{\prod\limits_{l=1}^{k-1}\upalpha_l(\bm{x}_{1:l})}{E[\prod\limits^{k-1}_{l=1}\upalpha_l(\bm{x}_{1:l}) ]} 
\notag\\
&= \frac{\prod\limits_{l=1}^k\upalpha_l(\bm{x}_{1:l})}{E[\prod\limits^{k-1}_{l=1}\upalpha_l(\bm{x}_{1:l}) ]}.\notag
\end{align}

Then,
\begin{align}
\bar{\upalpha}_k \stackrel{p}{\longrightarrow} E\left\{\frac{\prod\limits_{l=1}^k\upalpha_l(\bm{x}_{1:l})}{E[\prod\limits^{k-1}_{l=1}\upalpha_l(\bm{x}_{1:l}) ]}\right\}= \frac{E[\prod\limits^k_{l=1}\upalpha_l(\bm{x}_{1:l}) ]}{E[\prod\limits^{k-1}_{l=1}\upalpha_l(\bm{x}_{1:l}) ]}. \notag
\end{align}

Therefore, 
\begin{align}\notag
\frac{H^i_k}{g^*_k(\bm{x}^i_{1:k})} &= \frac{\bar{\upalpha}_1 \cdots \bar{\upalpha}_k}{\prod\limits^k_{l=1}\upalpha_l(\bm{x}_{1:l})} \frac{\prod\limits_{l=1}^k\upalpha_l(\bm{x}_{1:l})}{E[\prod\limits^k_{l=1}\upalpha_l(\bm{x}_{1:l}) ]}\\ & = E[\prod\limits^k_{l=1}\upalpha_l(\bm{x}_{1:l}) ]^{-1}\bar{\upalpha}_1 \cdots \bar{\upalpha}_k \stackrel{p}{\longrightarrow} 1\ \ as\ m\rightarrow 0. \label{eq60}
\end{align}

Similarly, we have
\begin{align}
\frac{\tilde{H}^i_k}{g^*_k(\tilde{\bm{x}}^i_{1:k})}\stackrel{p}{\longrightarrow} 1.\label{eq61}
\end{align}

Applying Lemma 1 (i) to $|\bm{G}(\cdot)|\mathbb{1}_{\{|\bm{G}(\cdot)| > M\}}$  for $M > 0$,
\begin{align}
m^{-1}\sum\limits^m_{i=1}|\bm{G}(\tilde{\bm{x}}^i_{1:k})|\mathbb{1}_{\{|\bm{G}(\tilde{\bm{x}}^i_{1:k})| > \upepsilon \sqrt{m}\}} \leq \notag\\
\ \ \ \ \ \ \ \ m^{-1}\sum\limits^m_{i=1}|\bm{G}(\tilde{\bm{x}}^i_{1:k})|\mathbb{1}_{\{|\bm{G}(\tilde{\bm{x}}^i_{1:k})| > M\}}.
\end{align}

For arbitrary M, when m large enough. 
\begin{align}
m^{-1}\sum\limits^m_{i=1}|\bm{G}(\tilde{\bm{x}}^i_{1:k})|\mathbb{1}_{\{|\bm{G}(\tilde{\bm{x}}^i_{1:k})| > M\}} \\
\stackrel{p}{\longrightarrow} E[|\bm{G}(\bm{x}_{1:k})|\mathbb{1}_{\{|\bm{G}(\bm{x}_{1:k})| > M\}}].\notag
\end{align}

As $M \rightarrow \infty$, $E[|\bm{G}(\bm{x}_{1:k})|\mathbb{1}_{\{|\bm{G}(\bm{x}_{1:k})| > M\}}] \rightarrow 0$. Therefore,
\begin{align}
m^{-1}\sum\limits^m_{i=1}|\bm{G}(\tilde{\bm{x}}^i_{1:k})|\mathbb{1}_{\{|\bm{G}(\tilde{\bm{x}}^i_{1:k})| > \upepsilon \sqrt{m}\}}  \stackrel{p}{\longrightarrow} 0. \label{eq64}
\end{align}

\subsection{Unconditional Distribution} 

In this part, we show that $ \sqrt{m}(\hat{\bm{x}}^*_{T}-E (\bm{x}_{T} |\bm{z}_{1:T} )) $ is asymptotically normal by induction.

By equation \eqref{eq76}, we have that
\begin{equation}\label{eq77}
E \left [ \left. e^{iu^{T} \sum_{j=1}^{m} M_{k}^{j}/\sqrt{m}}  \right | \mathcal{F}_{k-1}        \right ]\stackrel{p}{\rightarrow} e^{ -\frac{1}{2} u^{T} \bm{\Sigma}_{k}   }. 
\end{equation}

Then, our goal is to show unconditional asymptotic distribution of $ \sqrt{m}(\hat{\bm{x}}^*_{T}-E (\bm{x}_{T} |\bm{z}_{1:T} )) $ converges to the same characteristic function as that of normal distribution.
\begin{equation}\label{eq78}
E \left [  e^{iu^{T} \sqrt{m}(\hat{\bm{x}}^*_{k}-E (\bm{x}_{k} |\bm{z}_{1:k} )) }\right ]\stackrel{p}{\rightarrow} e^{ -\frac{1}{2} u^{T} \bm{\Sigma} u  } .
\end{equation}

Specifically, we want to prove that:
\begin{align}\label{eq79}
&E \left [  e^{iu^{T} \sqrt{m}(\hat{\bm{x}}^*_{k}-E (\bm{x}_{k} |\bm{z}_{1:k} )) }\right ]\notag\\
&= E \left [  e^{iu^{T} \sum_{k=1}^{2T-2} ( \sum_{j=1}^{m} M^{j}_{k} / \sqrt{m} ) } E \left( \left. e^{ \sum_{j=1}^{m}  M^{j}_{2T-1} / \sqrt{m}}  \right | \mathcal{F}_{2T-2} \right) \right ] \notag\\
&= e^{ -\frac{1}{2} u^{T} \bm{\Sigma}_{2T-1} u} E \left( e^{iu^{T} \sum_{k=1}^{2T-2} ( \sum_{j=1}^{m} M^{j}_{k} / \sqrt{m} ) }  \right) + o(1).
\end{align}

First, let us check that when $T = 1$, \eqref{eq79} is automatically satisfied since
\begin{equation} \notag
E \left [  e^{iu^{T} \sqrt{m}(\hat{\bm{x}}^*_{1}-E (\bm{x}_{1} |\bm{z}_{1} )) }\right ] = e^{ -\frac{1}{2} u^{T} \bm{\Sigma}_{1} u} E \left( e^{iu^{T} \times 0 }  \right) + o(1).
\end{equation}

Next, assuming that when $T=K$, it is satisfied:
\begin{equation}\notag
\begin{split}
e^{ -\frac{1}{2} u^{T} \bm{\Sigma}_{2K-1} u} E \left( e^{iu^{T} \sum_{k=1}^{2K-2} ( \sum_{j=1}^{m} M_{k}^{j}/\sqrt{m} ) }  \right) + o(1) \\
= e^{ -\frac{1}{2} u^{T} { \sum_{k=1}^{2K-1} } \bm{\Sigma}_{k} u }.
\end{split}
\end{equation}

Then, at $T=K+1$, it follows that:
\begin{align}\notag
&E \left [  e^{iu^{T} \sqrt{m}(\hat{\bm{x}}^*_{K+1}-E (\bm{x}_{k+1} |\bm{z}_{1:k+1} )) }\right ]\notag\\
&= e^{ -\frac{1}{2} u^{T} \bm{\Sigma}_{2K+1} u} E \left[ e^{iu^{T} \sum_{k=1}^{2K} ( \sum_{j=1}^{m} M_{k}^{j} / \sqrt{m} ) }  \right] + o(1) \notag\\
&= e^{ -\frac{1}{2} u^{T} \bm{\Sigma}_{2K+1} u} \times \notag\\
&\ \ \ \ E \left[ e^{iu^{T} \sum_{k=1}^{2K-2} ( \frac{\sum_{j=1}^{m} M_{k}^{j}}{\sqrt{m}}) + iu^{T} \left(\frac{\sum_{j=1}^{m} M^{j}_{2K-1}}{ \sqrt{m}} + \frac{\sum_{j=1}^{m} M^{j}_{2K}}{ \sqrt{m}} \right) }  \right]\notag \\  
&\quad \quad \quad \quad \quad \quad \ \ \ \ \   \quad  \quad  \quad  \quad  \quad  \quad  \quad \quad  \quad + o(1)\notag \\
&= e^{ -\frac{1}{2} u^{T} \left( \bm{\Sigma}_{2K-1} + \bm{\Sigma}_{2K} + \bm{\Sigma}_{2K+1} \right) u} e^{ -\frac{1}{2} u^{T} { \sum_{k=1}^{2K-2} } \bm{\Sigma}_k u } + o(1) \notag\\
&= e^{ -\frac{1}{2} u^{T} \left[ { \sum_{k=1}^{2K+1} } \bm{\Sigma}_{k}\right] u } + o(1).\notag
\end{align}

From above, we have shown that \eqref{eq78} holds. Therefore, the unconditional distribution of $ \sqrt{m}(\hat{\bm{x}}^*_{T}-E (\bm{x}_{T} |\bm{z}_{1:T} )) $ is asymptotically normal with mean zero and covariance matrix $ \bm{\Sigma}$. Since we have shown that, $\hat{\bm{x}}^*_{T} \stackrel{p}{\rightarrow}  \hat{\bm{x}}_T $, we can conclude that $ \sqrt{m}(\hat{\bm{x}}_{T}-\bm{x}_{T}) \stackrel{\text{dist}}{\longrightarrow} N (\bm{0}, \bm{\Sigma})$.









\end{document}